\documentclass[prl,twocolumn,showpacs,preprintnumbers,amsmath,amssymb]{revtex4}

\usepackage{graphicx}
\usepackage{dcolumn}
\usepackage{bm}

\begin{document}

\title{Proximity Effect and Multiple Andreev Reflections in Diffusive
Superconductor-Normal-Metal-Superconductor Junctions}

\author {J.C. Cuevas$^{1,2}$, J. Hammer$^1$, J. Kopu$^{1,3}$, J.K. Viljas$^1$,
and M. Eschrig$^1$}

\affiliation{$^1$Institut f\"ur Theoretische Festk\"orperphysik,
Universit\"at Karlsruhe, 76128 Karlsruhe, Germany}
\affiliation{$^2$Departamento de F\'{\i}sica Te\'orica de la Materia 
Condensada, Universidad Aut\'onoma de Madrid, 28049-Madrid, Spain}
\affiliation{$^3$Low Temperature Laboratory, Helsinki University of
Technology, P.O.Box 2200, FIN-02015 HUT, Finland}

\date{\today}

\begin{abstract}
 We present a theory of the current-voltage characteristics in diffusive 
 superconductor-normal-metal-superconductor junctions. By solving the
 time-dependent Usadel equations we are able to describe the phase-coherent
 transport for arbitrary length of the normal wire. We show how the interplay
 between proximity effect and multiple Andreev reflections gives rise to a 
 rich subgap structure in the conductance and how it is revealed in the 
 non-equilibrium distribution function. 
\end{abstract}


\maketitle

Proximity effect is the modification of the properties of a normal metal (N) 
in contact with a superconductor (S) and it has been extensively studied in 
diffusive hybrid nanostructures~\cite{Pannetier2000}. Both 
equilibrium~\cite{Gueron1996} and transport properties~\cite{Courtois1999,Dubos2001a}
of diffusive SN systems are now well understood in the framework of the Usadel
equations~\cite{Usadel1970}. The transport through an SN interface is mediated
by Andreev reflection, where an electron coming from N with energy $\epsilon$ 
below the superconducting gap $\Delta$ is converted into
a reflected hole, thus transferring a Cooper pair to the S electrode.
The time-reversed states involved in this process are coherent over a distance 
$L_C = \mbox{min} (\sqrt{\hbar D /\epsilon}, L_{\phi})$, where $D$ is the diffusion
constant of N and $L_{\phi}$ is the phase coherence length. 

In an SNS junction the transport at finite bias is dominated by multiple Andreev
reflections (MARs). Here, successive Andreev reflections at both S 
electrodes lead to a progressive rise of the quasiparticle energy. 
This process continues until the quasiparticle energy exceeds the gap energy.
A microscopic theory of MARs has emerged recently and has
been shown to describe quantitatively the current-voltage (I-V)
characteristics~\cite{Scheer1997}, the noise~\cite{Cron2001} and the
supercurrent~\cite{Goffman2000} in atomic point contacts. 

The interplay between proximity effect and MARs in diffusive SNS systems gives
rise to a rich variety of physical phenomena. For instance, the conductance 
exhibits a very peculiar subgap 
structure~\cite{Kutchinsky1997,Taboryski1999,Hoss2000}. This interplay is also
revealed in the noise~\cite{Hoss2000,Hoffmann2004} or in the Shapiro 
steps~\cite{Dubos2001b}. The understanding of these experiments is a basic 
open problem in mesoscopic superconductivity.

The theory of dissipative transport in diffusive SNS junctions has
mainly been developed in two limits. The first one is the incoherent
regime, when the normal metal length $L > L_C$. In this case there is no proximity
effect and the transport can be described in terms of a semiclassical kinetic
equation for the distribution function~\cite{Bezuglyi2000}. This function was 
actually measured in Ref.~\onlinecite{Pierre2001}, and it was shown to exhibit a 
step-like structure, which is a manifestation of MARs. On the other hand, 
in short SNS junctions, when $L \ll \xi=\sqrt{\hbar D/\Delta}$, where $\xi$ is
the superconducting coherence length, i.e. when the Thouless energy
$\epsilon_T = \hbar D/L^2$ exceeds $\Delta$, the MARs are fully coherent. 
In this regime the transport can be described by averaging the single-channel
point-contact results with the bimodal transmission distribution for 
diffusive systems~\cite{Bardas1997}. For the intermediate regime 
$\xi<L<L_{\phi}$, when the interplay between proximity effect and MARs takes 
place~\cite{Samuelsson2002}, there is no satisfactory theory so far.

In this Letter we study the phase-coherent transport in diffusive
voltage-biased SNS systems. We have solved the time-dependent Usadel equations,
which allows us to calculate the I-V characteristics for the whole range of
lengths from the short junction limit ($L \ll \xi$) to the incoherent regime
($L \gg \xi$). We show that the interplay between proximity effect and MARs
gives rise to a rich structure in the conductance in good agreement 
with existing experiments. We also predict how this interplay is manifested
in the quasiparticle distribution function.

We consider an SNS junction, where N is a diffusive normal metal of length 
$L < L_{\phi}$ coupled to two identical superconducting reservoirs. We assume
the SN interfaces to be fully transparent and neglect the suppression of the
pair potential in the S leads near the interfaces. Our goal
is the calculation of the current when a constant voltage $V$ is applied. For
this purpose we use the quasiclassical theory of superconductivity for diffusive
systems~\cite{Usadel1970}. This theory is formulated in terms of 
momentum averaged Green functions ${\bf \check G}({\bf R}, t,
t^{\prime})$ which depend on position ${\bf R}$ and two time
arguments. These propagators are $2\times 2$ matrices
in Keldysh space ($\check \; $), where each entry is a $2\times 2$ matrix in 
electron-hole space ($\hat \; $):
\begin{equation}
\label{keldysh-space}
{\bf \check G} = \left( \begin{array}{cc}
\hat G^R & \hat G^K \\
   0     & \hat G^A
\end{array} \right); \hspace{5mm} 
\hat G^{R} = \left( \begin{array}{cc}
{\cal G}^{R} & {\cal F}^{R} \\
\tilde {\cal F}^{R}  & \tilde {\cal G}^{R}
   \end{array} \right) .
\end{equation}
\noindent
The Green functions for the left (l) and right (r) leads can be written 
as ${\bf \check G}_{l,r}(t,t^{\prime}) = e^{-i \mu_{l,r}t \hat \tau_3/\hbar} 
{\bf \check G}_0(t-t^{\prime}) e^{i \mu_{l,r}t^\prime \hat \tau_3/ \hbar}$, 
where for the chemical potentials we use $\mu_l=0$, $\mu_r=eV$.
Here, ${\bf \check G}_0(t)$ is the equilibrium bulk Green function
of a BCS superconductor. We transform to energy representation, in which the
propagator ${\bf \check G}({\bf R},\epsilon,\epsilon^{\prime})$ depends on two 
energy arguments. It satisfies the non-stationary Usadel equation, 
which in the N region reads
\begin{equation}
\label{usadel-eq}
 \frac{\hbar D}{\pi} \nabla \left( {\bf \check G} \circ \nabla {\bf \check G} 
 \right) + \epsilon \hat{\tau}_3  {\bf \check G} - {\bf \check G} \hat{\tau}_3 
 \epsilon^{\prime} = 0 ,
\end{equation}

\noindent
where $\hat{\tau}_3$ is the Pauli matrix in electron-hole space. The convolution 
product $\circ$ is defined as $({\bf \check A} \circ {\bf \check B})
(\epsilon,\epsilon^{\prime}) = \int d\epsilon_1 \; {\bf \check A}(\epsilon,
\epsilon_1) {\bf \check B}(\epsilon_1, \epsilon^{\prime})$. Equation 
(\ref{usadel-eq}) is supplemented by the normalization condition ${\bf \check G}
\circ {\bf \check G} = -\pi^2 \delta(\epsilon - \epsilon^{\prime}){\bf \check 1}$. 

In order to solve the Usadel equation it is convenient to use the time-dependent
Riccati parametrization~\cite{Eschrig2000}. In this method the Green functions 
are parametrized in terms of scalar retarded (R) and advanced (A)
coherence functions $\gamma^{R,A}({\bf R},\epsilon,\epsilon^{\prime})$ 
and $\tilde \gamma^{R,A}({\bf R}, \epsilon,\epsilon^{\prime})$, and two 
distribution functions $x({\bf R},\epsilon, \epsilon^{\prime})$ and 
$\tilde x({\bf R},\epsilon, \epsilon^{\prime})$ as follows
\begin{eqnarray}
 \label{riccati}
 \hat G^{R} & = & - i \pi \; \hat N^{R} \circ \hat M^{R} ,
 \quad \hat G^{A}  =  i \pi \; \hat M^{A} \circ \hat N^{A} \nonumber \\
 \hat G^K &=& - 2 \pi i \; \hat N^R  \circ \hat M^K \circ \hat N^A
\end{eqnarray}

\noindent
with the abbreviations
\begin{eqnarray}
 \hat M^{R,A} & = & 
 \left( \begin{array}{cc}
 1 - \gamma^{R,A} \circ \tilde \gamma^{R,A} & 2 \gamma^{R,A} \\
 2 \tilde \gamma^{R,A} & \tilde \gamma^{R,A} \circ \gamma^{R,A} -1 
 \end{array} \right) \nonumber \\
 \hat M^K & = & 
\left( \begin{array}{cc}
 x + \gamma^R \circ \tilde x \circ \tilde \gamma^A & \quad
 x \circ \gamma^A - \gamma^R \circ \tilde x \\
 \tilde \gamma^R \circ x - \tilde x \circ \tilde \gamma^A &\quad
 \tilde x + \tilde \gamma^R \circ x \circ \gamma^A 
 \end{array} \right) \nonumber \\
\hat N^{R,A} &=&  \left( 
\begin{array}{cc}
1 + \gamma^{R,A} \circ \tilde \gamma^{R,A} & 0 \\
0 & 1 + \tilde \gamma^{R,A} \circ \gamma^{R,A}  
\end{array} \right)^{-1} \nonumber ,
\end{eqnarray}
\noindent
where the inverse is defined via the $\circ$-operation. Using fundamental 
symmetries, all the Green functions can be obtained from $\gamma^R$ and $x$. 
The transport equations for these functions in the N wire are~\cite{Eschrig2004}
\begin{eqnarray}
\label{g-eq}
\partial^2_z \gamma^R + (\partial_z \gamma^R) \circ \frac{\tilde {\cal F}^R}{i\pi}
 \circ (\partial_z \gamma^R) 
&=& \frac{{\cal E}\circ \gamma^R + \gamma^R \circ {\cal E}}{i\epsilon_T} \qquad
\\
\label{x-eq}
\partial^2_z x - (\partial_z \gamma^R) \circ \frac{\tilde {\cal G}^K}{i\pi}
 \circ (\partial_z \tilde \gamma^A) &+& (\partial_z \gamma^R) \circ 
 \frac{\tilde {\cal F}^R}{i\pi} \circ (\partial_z x) \nonumber \\
\qquad - (\partial_z x) \circ \frac{{\cal F}^A}{i\pi} \circ
 (\partial_z \tilde \gamma^A) 
 &=& \frac{{\cal E}\circ x - x \circ {\cal E}}{i\epsilon_T}, \qquad
\end{eqnarray}
\noindent
where ${\cal E}(\epsilon, \epsilon')\equiv \epsilon \cdot 
\delta(\epsilon-\epsilon')$. Here, $0<z<1$ is the dimensionless
coordinate which describes the position along the N wire. 
The expressions for $\tilde {\cal F}^R$, ${\cal F}^A$, and $\tilde {\cal G}^K$ 
are obtained by comparing Eq.~(\ref{keldysh-space}) with Eq.~(\ref{riccati}). 
A solution of these equations can be found using the Ansatz: 
\begin{equation}
\gamma^R(\epsilon,\epsilon^{\prime}) = \sum_m \gamma^R_{0,m}(\epsilon)
\delta(\epsilon - \epsilon^{\prime} + \omega_{m})
\label{ansatz}
\end{equation}
\noindent
where $\omega_m\equiv 2meV$. Other Fourier components are defined via
$\gamma^R_{n,m}(\epsilon)=\gamma^R_{0,m-n}(\epsilon+\omega_n)$.
Using the above Ansatz for any two functions $A$ 
and $B$, the Fourier components of $(A\circ B) (\epsilon , \epsilon')$
are given by $[A\circ B]_{n,m}(\epsilon ) =
\sum_l A_{n,l} (\epsilon ) B_{l,m} (\epsilon )$.
The equations for the Fourier components $\gamma^R_{n,m}(\epsilon )$ and
$x_{n,m}(\epsilon )$ are the same as Eqs.~(\ref{g-eq},\ref{x-eq}),
where the $\circ$-product denotes now a matrix product in the Fourier
indices and ${\cal E}_{n,m}(\epsilon )= (\epsilon + \omega_n) \cdot \delta_{n,m}$.
The boundary conditions for Eqs.~(\ref{g-eq},\ref{x-eq}) can be easily deduced
from the expressions of the bulk Green functions \cite{bound}. 
In general, these equations have to be solved numerically. 

\begin{figure}[t]
\begin{center}
\includegraphics[width=\columnwidth,clip]{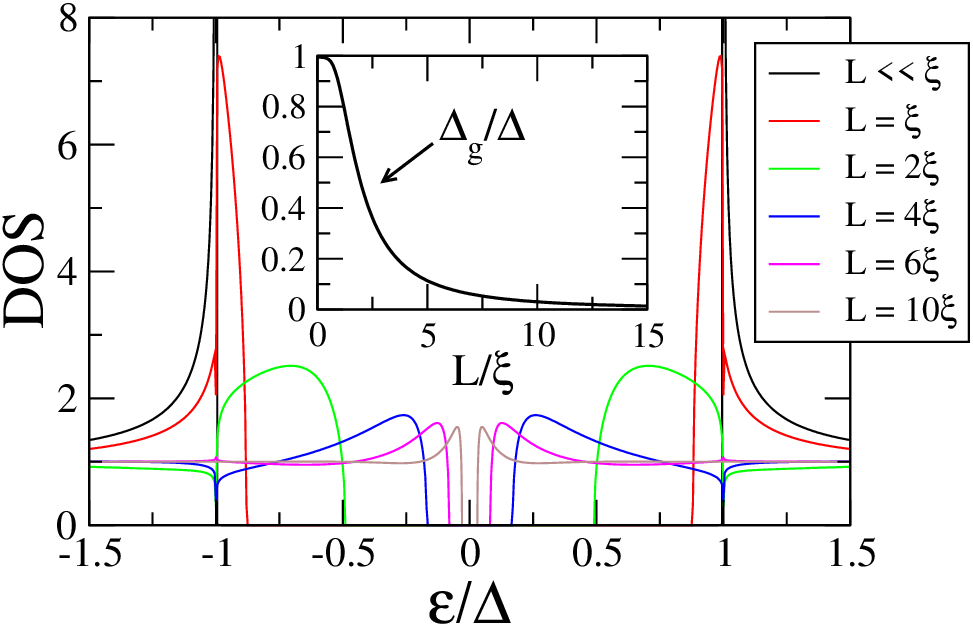}
\caption{\label{dos} (Color online) Normalized density of states in the 
middle of the wire as a function of energy for different wire lengths.
Inset: minigap $\Delta_g$ as a function of the wire length.}
\end{center}
\end{figure}

The above Ansatz leads to a current of the form $I(t) = \sum_m I_m \exp(i m 
\omega_J t)$, where $\omega_J = 2eV/\hbar$ is the Josephson frequency, and the 
current components can be written in terms of the Fourier components of the
Green functions, ${\bf \check G}_{n,m}(\epsilon)$. We concentrate here on 
the analysis of the dc current $I_0$, which we shall denote as $I$. It reads
\begin{equation}
I = \frac{G_N}{8\pi^2 e} \int_0^{2eV} d\epsilon \; \sum_{m} \mbox{Tr} 
\left\{ \hat \tau_3 \left[ {\bf \check G}\circ \partial_z {\bf \check G}
\right]^K_{m,m} (\epsilon ) \right\}, 
\end{equation}
\noindent
where $G_N$ is the normal state conductance. Next we express $I$ in
terms of the distribution function. It is possible to relate the component 
$\hat M^K_{11}$ to the electron distribution function 
$f({\bf R},\epsilon,\epsilon^{\prime})$ via the relation
$2f = 1-\sum^{\infty}_{n=0} (\gamma^R \circ \tilde \gamma^R)^n \circ 
\hat M^K_{11} \circ (\gamma^A \circ \tilde \gamma^A)^n$. Combining this with 
fundamental symmetries of the Green functions we can write the current as
\begin{equation}
\label{dc-current}
I = \frac{G_N}{e} \int_0^{2eV} d\epsilon \sum_{m}\Big\{ [\mathcal D 
\circ \partial_z f]_{m,m} -\mbox{Re} \left[ \mathcal S\circ f \right]_{m,m} \Big\},
\end{equation}
\noindent
where $\mathcal D = 1/2 +
({\cal G}^A \circ {\cal G}^R - {\cal F}^A \circ \tilde {\cal F}^R)/2\pi^2$ and 
$\mathcal S = ({\cal G}^R \circ \partial_z {\cal G}^R + {\cal F}^R \circ 
\partial_z \tilde {\cal F}^R)/\pi^2$. Here, $\mathcal D$ describes the 
renormalization of the diffusion constant and $\mathcal S $ describes the
spectral supercurrent.  

\begin{figure}[t]
\begin{center}
\includegraphics[width=\columnwidth,clip]{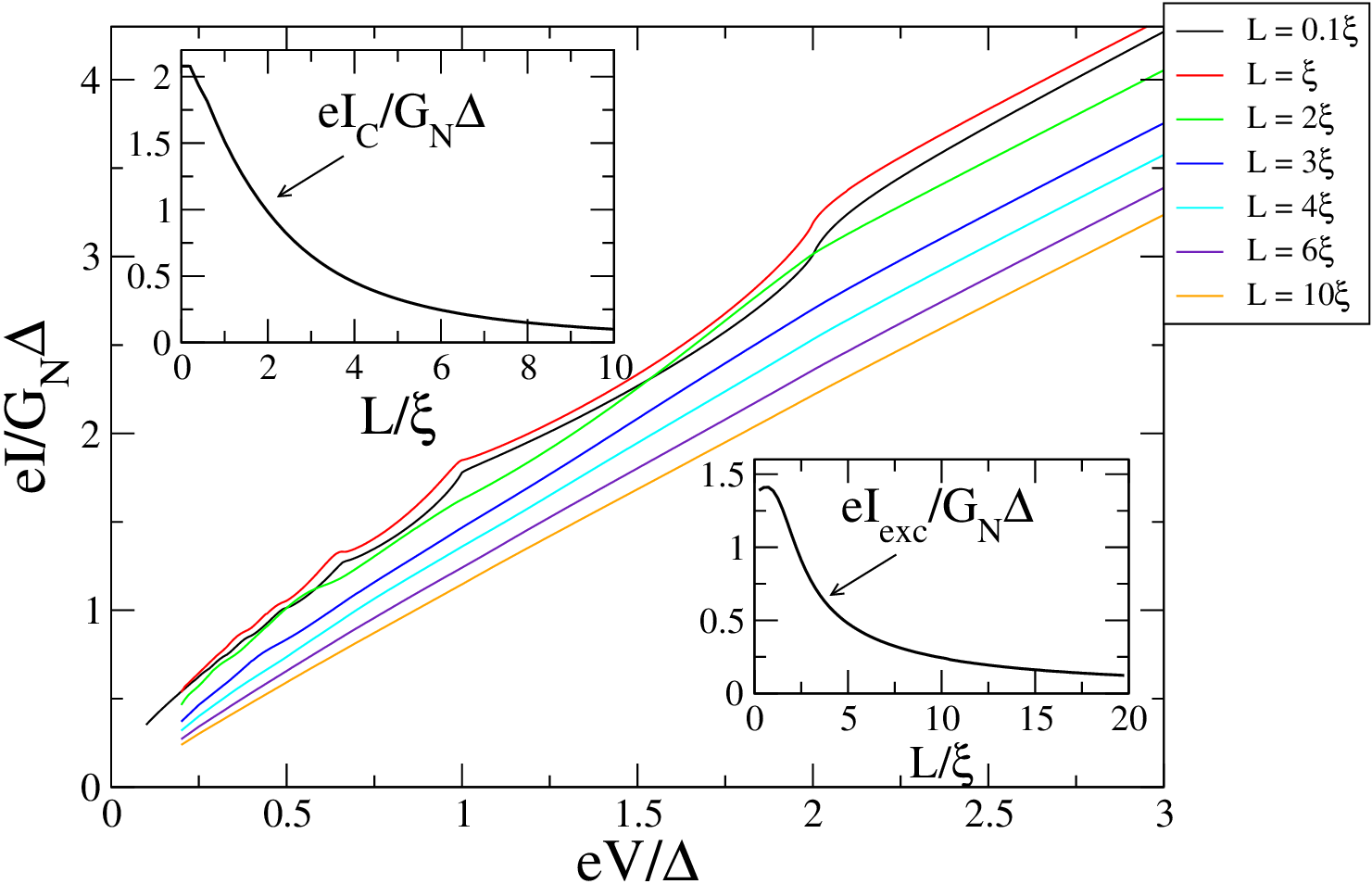}
\caption{\label{iv} (Color online) Zero-temperature dc current as a function of
the voltage for different wire lengths $L$. Upper inset: zero
temperature critical current as a function of $L$. Lower inset: excess 
current as a function of $L$.}
\end{center}
\end{figure}

It is instructive to first have a look at the zero-bias density of states (DOS).
In Fig.~\ref{dos} we show the DOS in the middle of the wire for different 
values of $L$. The most prominent feature is the presence of a minigap $\Delta_g$
which scales with the wire length as shown in the inset of Fig.~\ref{dos}. 
In the limit of long wires we obtain $\Delta_g \approx 3.2 \epsilon_T$. It is
worth remarking that while the DOS changes along the wire, the minigap is 
independent of the position.

Let us now turn to the analysis of the I-V characteristics. Figure \ref{iv} shows 
the zero-temperature I-V curves for different wire lengths. The main features are: 
(i) for $L \le \xi$ there is a pronounced subharmonic gap structure
(SGS) at voltages $2\Delta/ne$ ($n$ integer). In particular, the curve $L=0.1\xi$
reproduces quantitatively the result of the short-junction 
limit~\cite{Bardas1997}. (ii) For $L > \xi$ the SGS is progressively washed out 
as $L$ increases. (iii) At high bias ($eV>>\Delta$) there is an excess current, 
which is defined as $I_{exc} = I - G_N V$. In the lower inset of Fig.~\ref{iv} 
we show $I_{exc}$ as a function of $L$. For $L \to 0$ we recover the result 
$eI_{exc}/(G_N\Delta) = \pi^2/4 -1$ of Ref.~\onlinecite{Bardas1997}. 
In the opposite limit of long junctions ($L \gg \xi$) Volkov 
\emph{et al.}~\cite{Volkov1993} found that $I_{exc}$ decays according to 
$eI_{exc}/(G_N\Delta) = 0.82 \xi /L$. We find that $I_{exc}$ can be fitted to 
$eI_{exc}/(G_N\Delta) = 2.47\xi/ L$ in the experimentally relevant range 
$3\xi < L < 20\xi$, a factor of three larger than in the limit of 
Ref.~\onlinecite{Volkov1993}. 

As seen in Fig.~\ref{iv}, it is numerically difficult to reach the zero-bias 
limit, because the dimension of the matrices in Eqs.~(\ref{g-eq}-\ref{ansatz})
scales with the voltage roughly as $(2\Delta/eV)^2$. However, the analysis of 
the low bias regime is not the most relevant. The SNS junctions usually have a
negligible capacitance and their I-V curves are hence nonhysteretic, exhibiting
a transition from a supercurrent to a voltage state at the critical current. In
the upper inset of Fig.~\ref{iv} we show the value of the zero-temperature 
critical current, $I_C$. As can be seen, the transition to the supercurrent branch
would take place at voltages which are accessible to our numerical solution. 

\begin{figure}[t]
\begin{center}
\includegraphics[width=\columnwidth,clip]{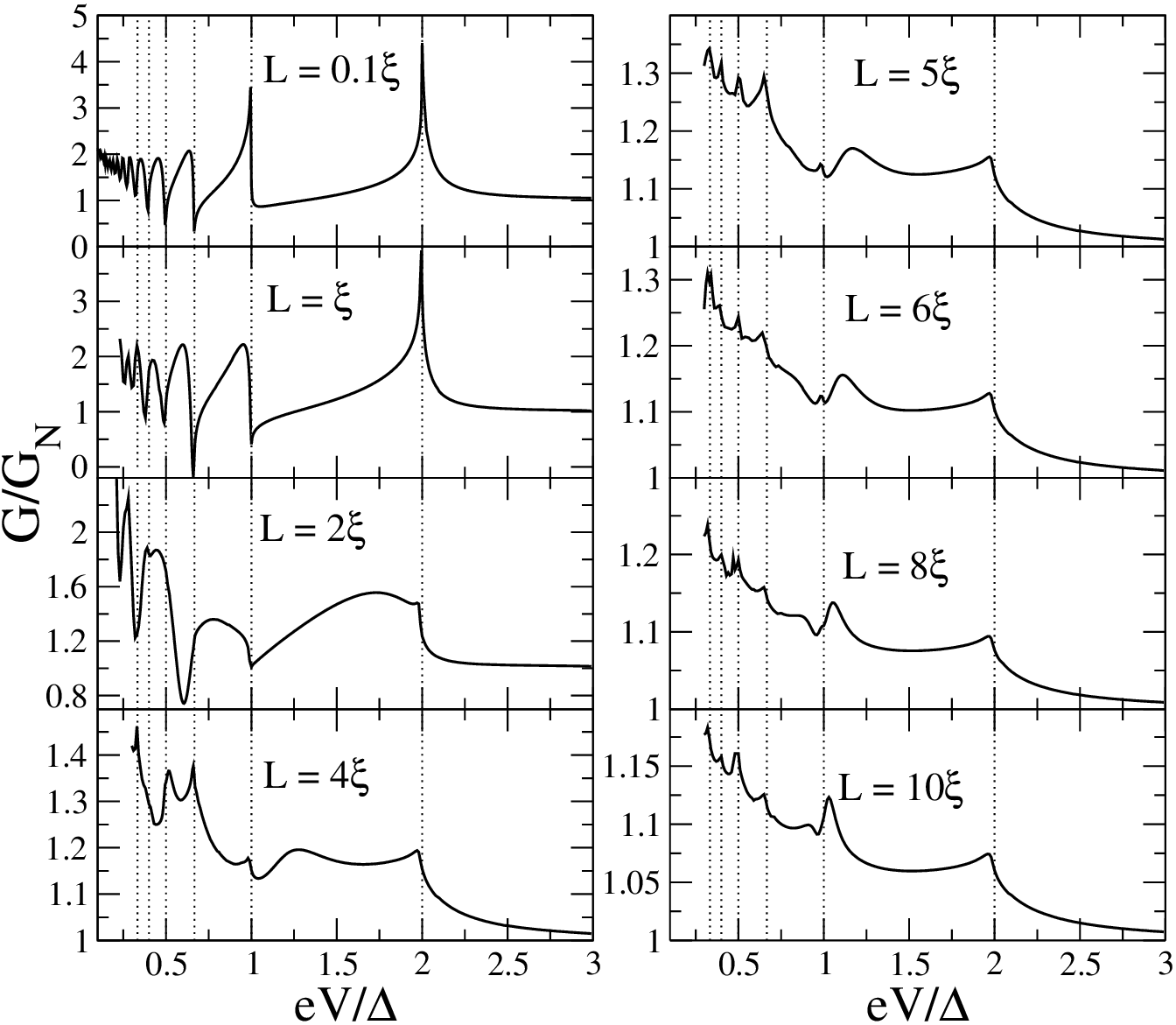}
\caption{\label{cond} Zero-temperature differential conductance as a 
function of the voltage for different wire lengths.
The vertical lines indicate the position of $eV=2\Delta/n$ with
$n=1,...,6$.} 
\end{center}
\end{figure}

The non-linearities in the I-Vs can more clearly be seen in the differential
conductance, $G=dI/dV$, which is shown in Fig.~\ref{cond}. Notice that
for $L \le \xi$ the SGS consists of a set of pronounced maxima at 
roughly $eV_n=2\Delta/n$ ($n$ integer). For $L > \xi$ the shape and height
of the maxima change drastically and new structure appears. 
For instance, the peak $n=2$ appears slightly below $eV=\Delta$ and it is 
accompanied by a much more pronounced maximum above $\Delta$. This maximum
shifts towards $\Delta$ as $L$ increases, until it merges with the 
peak at $\Delta$. The peak above $\Delta$ is a common feature of the experimental 
observations of the SGS~\cite{Kutchinsky1997,Taboryski1999,Hoss2000}. Notice
also that in the range $4\xi<L<10\xi$
the SGS is superimposed on a background that increases as the voltage decreases.
The correction in the conductance (as compared to $G_N$) diminishes 
as $L$ increases and, for instance, it reaches 15\%-20\% at low bias 
for $L=10\xi$. All these features are in qualitative agreement with the
experimental observations (see e.g. Fig.~5 in Ref.~\onlinecite{Hoss2000}).
However, a quantitative comparison would require us to extend our analysis 
to arbitrary transparency of the SN interfaces. 

The existence of the minigap $\Delta_g$ suggests that the SNS junction
should exhibit a subgap structure similar to that of contacts with two different
gaps~\cite{Samuelsson2002}. Indeed, the conductance maximum above $\Delta$ 
appears roughly at $eV \sim \Delta + \Delta_g$. However, one must be careful 
with this analogy. First, other features expected for asymmetric junctions, 
like a peak at $\Delta - \Delta_g$, are absent. Moreover, our calculations
show that as a function of temperature the conductance maxima shift
following the bulk gap and, in particular, the peak above $\Delta$ survives
even for temperatures higher than the minigap. The reason for the failure of
the above argument is that at finite bias the minigap does not survive, and
only a position-dependent pseudogap appears in the spectrum.

\begin{figure}[t]
\begin{center}
\includegraphics[width=0.98\columnwidth,clip]{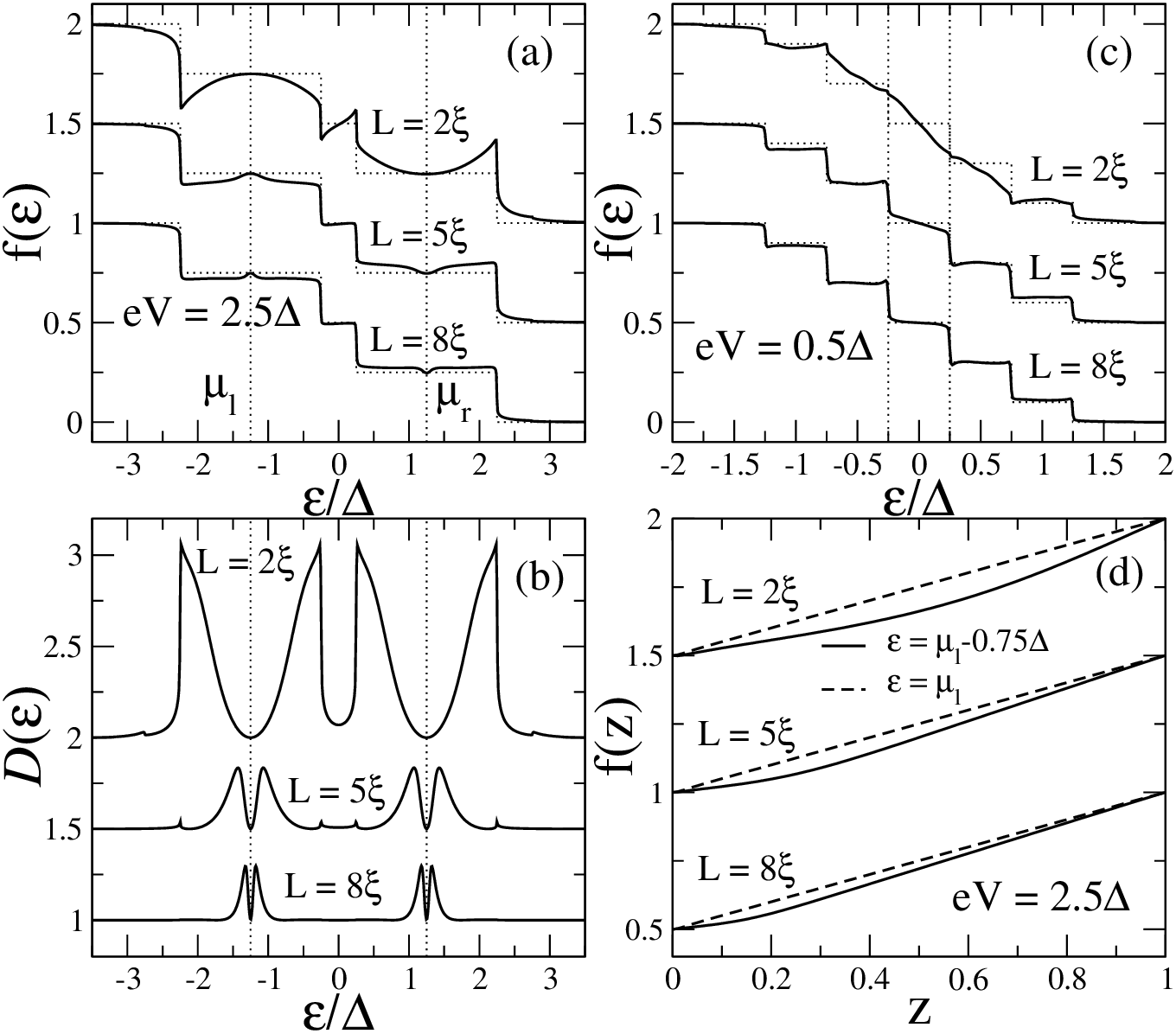}
\caption{\label{dist} (a) dc component of the distribution function 
in the middle of the wire ($z=1/2$) for three different wire lengths 
$L$ and $eV=2.5\Delta$.  The position of the chemical potentials are indicated
by vertical lines. The dotted lines are the results of the incoherent model 
(see text). (b) dc component of the renormalization of the diffusion constant 
for the cases shown in panel (a). (c) The same as panel (a) for $eV=0.5\Delta$. 
(d) Spatial variation along the wire of the distribution functions in panel 
(a) for fixed energies. The curves in (a)-(d) have been shifted by multiples 
of 0.5 for convenience.}
\end{center}
\end{figure}

In Fig.~\ref{dist} we present our results for the distribution function 
obtained for the intermediate regime, and compare with the results for the 
incoherent regime of Ref.~\onlinecite{Pierre2001}. We first summarize the 
main results of the incoherent model discussed in Ref.~\onlinecite{Pierre2001}. 
The basic assumptions are: (i) there is no proximity effect, and (ii)
the probability of Andreev reflection is 1 within the gap and 0 outside. 
As a result, the dc component of $f$ varies linearly with position $z$.
This model predicts a staircase pattern, and two examples are shown as dotted 
lines in Fig.~\ref{dist}(a,c), where the dc component of $f$ in the middle of 
the wire is plotted. Our full calculation shows pronounced deviations from 
this staircase pattern. An example can be seen in the energy 
region $-2.25\Delta < \epsilon < -0.25\Delta$ in Fig.~\ref{dist}(a),
where $eV=2.5\Delta$. In this region, the incoherent model predicts
$f = 3/4$, while the full calculation gives smaller values in the whole window, 
except exactly at $\epsilon = \mu_l$.

We explain these deviations in terms of the proximity effect. 
As shown in Fig.~\ref{dist}(d) for $eV=2.5 \Delta$, the proximity 
effect leads to a non-linear spatial variation of the distribution function 
$f$, except at the chemical potentials $\mu_{l,r}$. For $\epsilon=\mu_l$ 
the distribution function varies linearly from 0.5 to 1, as in the incoherent 
model. However, for $\epsilon=\mu_l-0.75\Delta $ the proximity effect near the 
left electrode leads to an effective shortening of the wire length.
This explains the negative deviation for the electron distribution function
in this energy range. For an energy near $\mu_r$ the deviation is positive, 
as the distribution function for this case (not shown here)
varies from 0 to 0.5 and the proximity effect takes place near $z=1$.

A related aspect is the renormalization of the  diffusion constant $\mathcal D$, 
the dc part of which is shown in Fig.~\ref{dist}(b). It reflects the enhancement
of the transmission with respect to the normal state due to the proximity effect. 
Notice the correlation between the structure in $\mathcal D$ and the deviations 
of $f$ from the staircase pattern.  In particular, at the
chemical potentials of the superconductors $\mathcal D = 1$, and $f$ 
adopts the values predicted by the incoherent model.

In summary, we have presented a theory of the I-V characteristics of diffusive 
SNS systems. We have studied how the interplay between coherent MARs and the
proximity effect is reflected in the conductance and in the distribution 
function.  Our main results are: (i) we reproduce an additional peak 
above $\Delta $ in the conductance, in agreement with experiments 
~\cite{Kutchinsky1997,Taboryski1999,Hoss2000}; 
(ii) we predict the distribution function for the intermediate regime, which
can be measured e.g. by the technique presented in Ref.~\onlinecite{Pierre2001}.
Our work paves the way for the study of other transport properties
in diffusive SNS systems~\cite{Hoffmann2004,Dubos2001b}.

We acknowledge fruitful discussions with D. Esteve, A. Levy Yeyati, 
A. Mart\'{\i}n-Rodero, H. Pothier, and G. Sch\"on.  This work has been 
financed by the Helmholtz Gemeinschaft and by the DFG within the CFN.



\vspace{-0.3cm}
\begin{thebibliography}{}
\vspace{-0.4cm}
\bibitem{Pannetier2000}
 B. Pannetier and H. Courtois, J. Low Temp. Phys. {\bf 188}, 599 (2000).

\bibitem{Gueron1996}
 S. Gu\'eron \emph{et al.}, Phys. Rev. Lett. {\bf 77}, 3025 (1996);
 N. Moussy \emph{et al.}, Europhys. Lett. {\bf 55}, 861 (2001). 

\bibitem{Courtois1999}
 H. Courtois \emph{et al.}, J. Low Temp. Phys. {\bf 116}, 187 (1999). 

\bibitem{Dubos2001a}
 P. Dubos \emph{et al.}, Phys. Rev. B {\bf 63}, 064502 (2001).

\bibitem{Usadel1970}
 K.D. Usadel, Phys. Rev. Lett. {\bf 25}, 507 (1970).

\bibitem{Scheer1997}
  E.~Scheer \emph{et al.}, Nature {\bf 394}, 154 (1998).

\bibitem{Cron2001}
  R.~Cron \emph{et al.}, Phys. Rev. Lett. {\bf 86}, 4104 (2001).

\bibitem{Goffman2000}
 M.F. Goffman \emph{et al.}, Phys. Rev. Lett. {\bf 85}, 170 (2000).

\bibitem{Kutchinsky1997}
 J. Kutchinsky \emph{et al.}, Phys. Rev. Lett. {\bf 78}, 931 (1997);
 J. Kutchinsky \emph{et al.}, Phys. Rev. B {\bf 56}, R2932 (1997).

\bibitem{Taboryski1999}
 R. Taboryski \emph{et al.}, Superlattices Microstruct. {\bf 25},
 829 (1999); J. Kutchinsky, Ph.D. thesis, Technical University of
 Denmark, 2001.

\bibitem{Hoss2000}
 T. Hoss \emph{et al.}, Phys. Rev. B {\bf 62}, 4079 (2000).

\bibitem{Hoffmann2004}
 C. Hoffmann \emph{et al.}, Phys. Rev. B {\bf 70}, 180503 (2004).

\bibitem{Dubos2001b}
 P. Dubos \emph{et al.}, Phys. Rev. Lett. {\bf 87}, 206801 (2001).

\bibitem{Bezuglyi2000}
 E.V. Bezuglyi \emph{et al.}, Phys. Rev. B {\bf 62}, 14439 (2000);
 E.V. Bezuglyi \emph{et al.}, Phys. Rev. B {\bf 63}, 100501 (2001);
 K.E. Nagaev, Phys. Rev. Lett. {\bf 86}, 3112 (2001);
 S. Pilgram and P. Samuelsson, Phys. Rev. Lett. {\bf 94}, 086806 (2005).

\bibitem{Pierre2001}
 F. Pierre \emph{et al.}, Phys. Rev. Lett. {\bf 86}, 1078 (2001).

\bibitem{Bardas1997}
 A. Bardas and D. Averin, Phys. Rev. B {\bf 56}, 8518 (1997);
 A.V. Zaitsev and D.V. Averin, Phys. Rev. Lett. {\bf 80}, 3602 (1998).

\bibitem{Samuelsson2002}
 P. Samuelsson \emph{et al.}, Phys. Rev. B {\bf 65}, 180514 (2002).

\bibitem{Eschrig2000}
 M. Eschrig, Phys. Rev. B {\bf 61}, 9061 (2000).

\bibitem{Eschrig2004}
 M. Eschrig \emph{et al.}, Adv. in Solid State Phys. {\bf 44}, 533 (2004).

\bibitem{bound}
$[x_{n,m}]_l(\epsilon) = x_0(\epsilon+\omega_n) \delta_{n,m}$,
$[\gamma^R_{n,m}]_l(\epsilon) = \gamma^R_0(\epsilon+\omega_n) \delta_{n,m}$, 
$[x_{n,m}]_r(\epsilon) = x_0(\epsilon+\omega_n+eV) \delta_{n,m}$, 
$[\gamma^R_{n,m}]_r(\epsilon) = \gamma^R_0(\epsilon+\omega_n+eV) \delta_{n,m-1}$, 
where $\gamma^R_0(\epsilon ) = -\Delta /\{\epsilon + 
i\sqrt{\Delta^2-(\epsilon+i0^+)^2} \}$ and
$x_0(\epsilon) = \tanh(\epsilon/2T) \cdot (1-|\gamma^R_0(\epsilon )|^2)$.

\bibitem{Volkov1993}
 A.F. Volkov \emph{et al.}, Physica C {\bf 210}, 21 (1993).

\end{thebibliography}
\end{document}